\documentclass[a4paper,11pt]{article}
\usepackage{pos}
\hypersetup{colorlinks,linkcolor={blue},citecolor={blue},urlcolor={blue}}
\title{Indirect detection of the QCD axion}

\newcommand{\TDLI}{\affiliation[a]{Tsung-Dao Lee Institute (TDLI), No.\ 1 Lisuo Road, 201210 Shanghai, China}}
\newcommand{\SJTU}{\affiliation[b]{School of Physics and Astronomy, Shanghai Jiao Tong University, 800 Dongchuan Road, 200240 Shanghai, China}}
\newcommand{\UVA}{\affiliation[c]{University of Virginia, Department of Astronomy, Charlottesville, VA, 22904, USA}}
\newcommand{\IFCA}{\affiliation[d]{Instituto de F\'isica de Cantabria (IFCA, UC-CSIC), Avenida de Los Castros s/n, 39005 Santander, Spain}}
\newcommand{\KCL}{\affiliation[e]{Department of Physics, King's College London, UK}}

\author*[a,b]{Luca Visinelli}
\TDLI\SJTU
\emailAdd{luca.visinelli@sjtu.edu.cn}

\author[c]{Bradley Johnson}
\UVA
\author[d]{Bradley J.\ Kavanagh}
\IFCA
\author[e]{\newline David J.\ E.\ Marsh}
\KCL
\author[c]{Jordan E.\ Shroyer}
\author[c]{Liam Walters}

\abstract{The QCD axion, originally proposed to solve the strong CP problem in QCD, is a prominent candidate for dark matter (DM). In the presence of strong magnetic fields, such as those around neutron stars, axions can theoretically convert into photons, producing detectable electromagnetic signals. This axion-photon coupling provides a unique experimental pathway to probe axions within a specific mass range. We investigate a novel observational approach using the Green Bank Telescope (GBT) to search for radio transients that could arise from interactions between neutron stars and dense DM clumps known as axion miniclusters. By observing the core of Andromeda with the VErsatile GBT Astronomical Spectrometer (VEGAS) and the X-band receiver (8 to 10 GHz), we achieve sensitivity to axions with masses in the range of 33 - 42\,$\mu$eV, with a mass resolution of $3.8 \times 10^{-4}\,\mu$eV. We detail our observational and analytical strategies developed to capture transient signals from axion-photon conversion, achieving an instrumental sensitivity of 2\,mJy per spectral channel. Despite our sensitivity threshold, no candidate signals exceeding the 5$\sigma$ level were identified. Future implementations will extend this search across additional spectral bands and refine the modeling used for the processes involved, strengthening the constraints on axion DM models. Based on Refs.~\cite{Walters:2024vaw, Edwards:2020afl, Kavanagh:2020gcy} and ongoing work.}

\FullConference{
2nd General Meeting of the COST Action COSMIC WISPers (CA21106) (COSMICWISPers2024)\\
3-6 September 2024\\
Istanbul (Turkey)\\
}

\begin{document}
\maketitle

\section{Introduction}
\label{sec:introduction}

The strong-CP problem can be addressed by introducing a new global symmetry into the Standard Model (SM) of particle physics, as proposed by Peccei and Quinn (PQ)~\cite{Peccei:1977hh, Peccei:1977ur}. This solution predicts the existence of the QCD axion, a pseudo-scalar particle~\cite{Weinberg:1977ma, Wilczek:1977pj}, which could serve as dark matter (DM) if produced through non-thermal processes that ensure the axion remains non-relativistic by the time of recombination~\cite{Abbott:1982af, Dine:1982ah, Preskill:1982cy}. An active search for the QCD axion and related axion-like particles is ongoing, see Refs.~\cite{Irastorza:2018dyq, DiLuzio:2020wdo, Chadha-Day:2021szb, Giannotti:2024xhx} for reviews, though detection is challenging due to the axion’s weak interactions with SM particles. Most experimental efforts focus on the axion-photon coupling, $g_{a\gamma\gamma}$, which modifies Maxwell’s equations in the presence of axions~\cite{Sikivie:1983ip, Sikivie:1985yu, Wilczek:1987mv, Krasnikov:1996bm, Visinelli:2013fia, Tercas:2018gxv, Visinelli:2018zif}. This coupling motivates various terrestrial experiments~\cite{Hagmann:1998cb, Asztalos:2001tf, Asztalos:2003px, Graham:2015ouw, Braine:2019fqb, Kenany:2016tta, Brubaker:2016ktl, TheMADMAXWorkingGroup:2016hpc, Majorovits:2016yvk, Alesini:2017ifp, Alesini:2019nzq, Kahn:2016aff, Ouellet:2018beu, Budker:2013hfa, Barbieri:2016vwg, Lee:2020cfj, Alesini:2022lnp, Alesini:2023qed, Gatti:2024mde, ADMX:2024pxg}, searches in helioscopes~\cite{OShea:2023gqn, OShea:2024jjw, Ruz:2024gkl}, and astrophysical searches for axion-photon conversion in regions with strong magnetic fields, such as those surrounding neutron stars (NSs)~\cite{Pshirkov:2007st, Huang:2018lxq, Hook:2018iia, Safdi:2018oeu, Leroy:2019ghm, Walters:2024vaw}.

The QCD axion's production history is closely tied to the thermal evolution of the early Universe. If PQ symmetry breaking occurs after inflation has ended, random fluctuations in the initial field conditions lead to the formation of self-gravitating clumps of axions around matter-radiation equality, known as axion miniclusters (AMCs)~\cite{Hogan:1988mp, Kolb:1993hw, Kolb:1993zz, Zurek:2006sy}. Simulations suggest that a substantial fraction of cold axions may reside within these bound structures, parameterized by $f_{\rm AMC}$, which can range from 1\% to nearly 100\%~\cite{Vaquero:2018tib, Buschmann:2019icd}. This fraction directly influences detection prospects: if most DM is bound in AMCs, Earth’s encounters with these structures become rare, limiting direct detection sensitivity~\cite{Sikivie:2006ni} (see Ref.~\cite{ADMX:2024pxg} for a recent laboratory search). AMC-like structures are also expected within inflationary scenarios~\cite{Hertzberg:2020hsz, Yin:2024xov}. As AMCs traverse the galactic halo, interactions with stars in the disk disrupt these structures. Repeated tidal stripping by stellar encounters gradually erodes miniclusters, altering their internal structure and spatial distribution over time~\cite{Berezinsky:2013fxa, Tinyakov:2015cgg}. A framework to quantify these effects on the AMC population was presented in Ref.~\cite{Kavanagh:2020gcy}, using a Monte Carlo approach to simulate AMC-stellar interactions. The model assumes a steady Milky Way structure post-formation and a simplified stellar population. Despite these assumptions, this analysis establishes a framework to assess the AMC survival rates and spatial distribution, with the associated numerical pipeline available at \href{https://github.com/bradkav/axion-miniclusters/}{github.com/bradkav/axion-miniclusters}. See also subsequent work on the topic in Refs.~\cite{Dandoy:2022prp, Shen:2022ltx, OHare:2023rtm, DSouza:2024flu}. Although this study focuses on miniclusters, AMCs may also host axion stars, another class of axionic objects formed through gravitational relaxation~\cite{Levkov:2018kau, Eggemeier:2019jsu, Chen:2020cef, Dmitriev:2023ipv}, with quantum pressure counteracting gravitational collapse (see, e.g.,~\cite{Visinelli:2017ooc, Schiappacasse:2017ham}). Additionally, AMC encounters with neutron star populations could produce transient radio signals observable as short bursts, as discussed in the companion analysis of Ref.~\cite{Edwards:2020afl}. The results predict a wide range of expected fluxes, from microjansky to several jansky for bright, detectable signals.

In recent years, efforts have focused on searching for axion-photon conversion signals from AMC-NS encounters across different radio frequency windows. This search was conducted using the VErsatile GBT Astronomical Spectrometer (VEGAS) receiver of the Green Bank Telescope (GBT), accounting for the environmental and astrophysical conditions. The findings, summarized in Ref.~\cite{Walters:2024vaw}, indicate that although individual encounters between AMCs and NSs are relatively rare, cumulative interactions across the NS population could produce detectable bursts of radio-frequency signals within a given time frame. These results align with earlier predictions~\cite{Edwards:2020afl} but suggest that the outcome is highly sensitive to the AMC distribution and tidal disruption effects. Several key parameters impacting detectability include the spatial distribution of AMCs, their internal density profiles, and unaccounted interactions between axion stars and compact stellar remnants. Regions within the galactic plane with a high density of axion stars were found to have a substantially higher probability of generating detectable signals compared to other parts of a galaxy. Notably, the presence of axion stars that survive tidal stripping could enhance detection prospects, as these remnants tend to cluster in localized regions. These aspects are discussed further below.

\section{Searching for transient events}

The axion field originates when the PQ symmetry spontaneously breaks. At much lower temperatures corresponding to the QCD phase transition, explicit symmetry breaking by QCD instantons drives the axion field to oscillate coherently around a CP-conserving minimum in a process known as vacuum realignment~\cite{Abbott:1982af, Dine:1982ah, Preskill:1982cy}. These oscillations store DM energy density within the axion condensate, determined by the initial misalignment angle $\theta_i$. This density is related to the axion mass, which is fixed by matching the current axion energy density to the observed DM abundance. The QCD axion’s production history is then directly linked to the thermal evolution of the early Universe~\cite{Visinelli:2009kt}. If the PQ symmetry breaks before inflation, $\theta_i$ is effectively homogeneous across the observable universe. However, if it breaks after inflation, fluctuations in the axion density become decoupled from cosmological expansion, and overdense regions collapse gravitationally around matter-radiation equality, forming self-gravitating structures known as AMCs~\cite{Hogan:1988mp, Kolb:1993zz, Kolb:1993hw, Zurek:2006sy}.

The AMC mass distribution follows an initial halo mass function that evolves as structure formation progresses, as shown in simulation results~\cite{Eggemeier:2019khm, Eggemeier:2022hqa}. The density within these gravitationally bound structures is determined by the initial overdensity, evaluated through numerical simulations~\cite{Vaquero:2018tib, Buschmann:2019icd, Buschmann:2021sdq}. The spatial distribution of AMCs within galaxies is typically modeled according to the DM density profile, such as the Navarro-Frenk-White (NFW) profile~\cite{Navarro:1995iw}. Tidal interactions with the mean galactic gravitational field and nearby stars further perturb both the AMC mass and spatial distribution, a process initially quantified in~\cite{Tinyakov:2015cgg, Dokuchaev:2017psd} and later refined using more detailed Monte Carlo approaches~\cite{Kavanagh:2020gcy, Dandoy:2022prp, Shen:2022ltx, OHare:2023rtm, DSouza:2024flu}. While a substantial portion of AMC mass may be stripped away, residual AMC cores are expected to persist, particularly in the outskirts of galaxies.

Ref.~\cite{Edwards:2020afl} explores potential radio signatures from axion-photon conversion during NS encounters with AMCs within the Milky Way. The study simulates a large sample of such encounters to predict distributions of fluxes, durations, and sky locations. Events are modeled by sampling key parameters: galactocentric radius, height above the Galactic plane, and azimuthal angle. The internal density profiles of AMCs are modeled with either a power-law or a NFW distribution. Radio emissions from AMC-NS encounters are characterized by narrow-band spectral profiles driven by axion velocity dispersion, with radio flux density estimated based on encounter distance and relevant physical parameters. The flux distribution, assuming isotropic emission and integrating over encounter durations, peaks in flux densities between $(10^{-6} \textrm{--} 10^2) \, \mu$Jy, with event durations typically spanning from days to several months. Notably, bright events exceeding the sensitivity threshold of radio telescopes like the Very Large Array are primarily produced by encounters with denser AMCs. For this analysis, a DM axion mass of 20\,$\mu$eV is assumed, though this parameter could vary over a broad range. The sky distribution of AMC-NS encounters in the Milky Way shows a concentration of events near the Galactic center, mirroring the NS spatial distribution. This spatial concentration could play a key role in detecting axion DM. The encounter rate is heavily influenced by the internal AMC density profile, with NFW profiles producing fewer bright events than PL profiles, although the latter has a higher encounter rate. These results suggest that current and forthcoming radio telescope capabilities are well-suited to detect such transient signals, presenting a promising path toward identifying axion DM in the near future.

\section{Results from GBT searches}

\begin{figure*}
\centering
\includegraphics[width=0.75\textwidth]{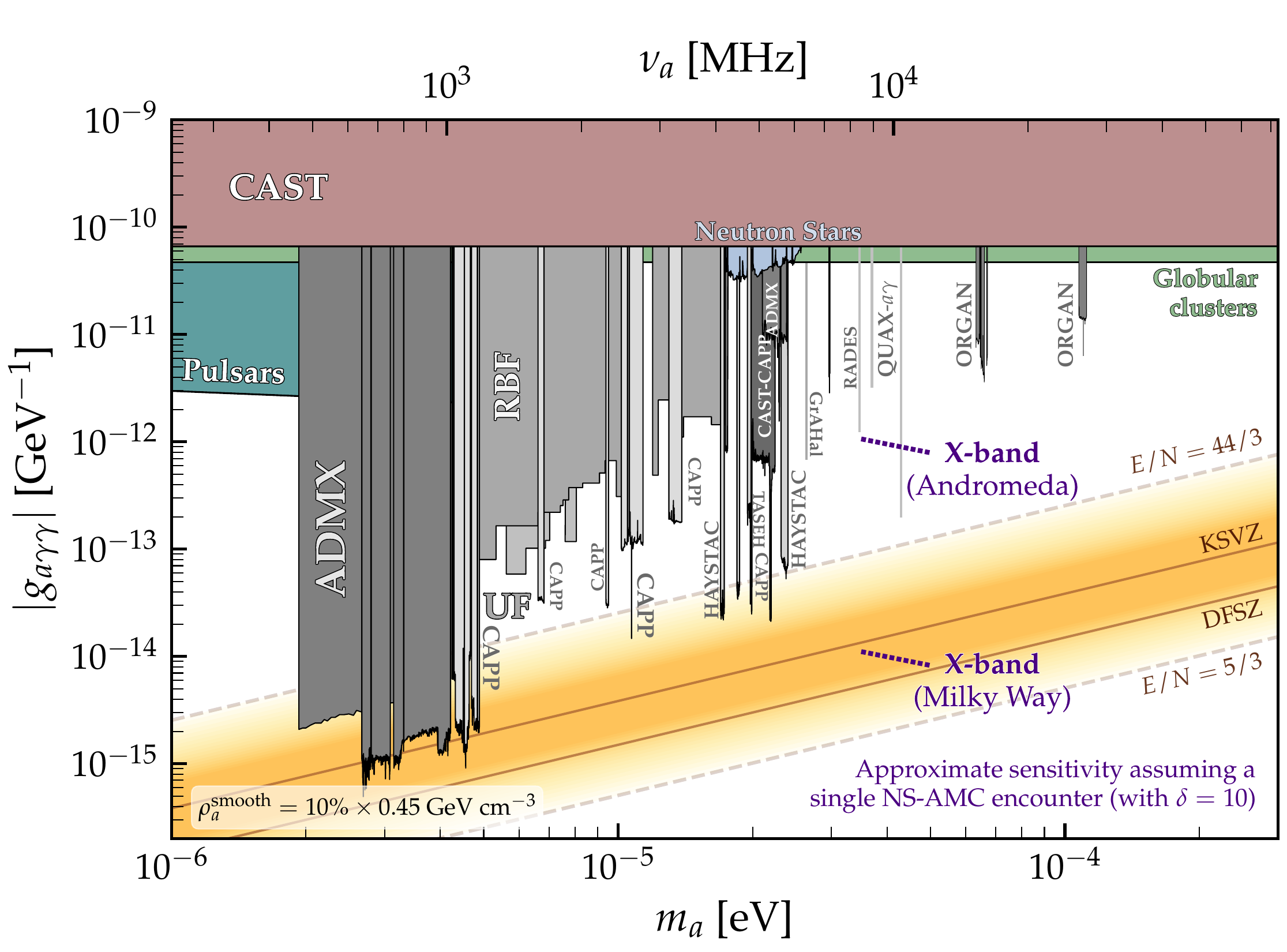}
\caption{Estimated sensitivity to axion-photon couplings assuming the observation of a single AMC-NS encounter, compared with the QCD axion model band (yellow) and a number of existing constraints from haloscopes, helioscopes, and astrophysical searches. Details are given in Ref.~\cite{Walters:2024vaw}. Figure adapted from \href{https://cajohare.github.io/AxionLimits/}{https://cajohare.github.io/AxionLimits/AxionLimits}.}
\label{fig:AxionLimits}
\end{figure*}

In August 2021, a proposal was made to observe the center of the Andromeda galaxy (M31) using the GBT in the X band, representing the first dedicated effort to search for radio transients potentially caused by AMC-NS collisions. The initial search, conducted in 2022 (GBT22A-067), consisted of four two-hour observation sessions. During this campaign, seven candidate signals were identified above the 5$\sigma$ detection threshold. However, none exhibited the characteristics expected of AMC-NS transients, leading to the conclusion that no detectable AMC-NS event occurred in M31 during the observation period~\cite{Walters:2024vaw}. Figure~\ref{fig:AxionLimits} illustrates the sensitivity reach for the axion-photon coupling parameter $g_{a\gamma\gamma}$ under specific assumptions regarding the overdensity parameters and the properties of the AMC involved.

Since the initial observations, theoretical modeling has advanced, now indicating a peak event rate near 3 GHz instead of the 10 GHz suggested in earlier studies~\cite{Edwards:2020afl, Witte:2021arp}. The updated model suggests that detectable events in the X-band are less frequent, requiring longer and broader observational windows to capture an AMC-NS collision. The event rate remains significant at particle masses below the expected range for cosmic axions that contribute to AMC formation~\cite{OHare:2021zrq, Gorghetto:2022ikz}. To investigate this, the search for AMC-NS collisions in M31 has continued, now focusing on the 3 GHz event-rate peak. In 2023, follow-up observations (GBT23A-245) were conducted using the GBT’s C-band receiver, covering frequencies from 4.0 to 8.0 GHz (16 to 33 $\mu$eV axion mass range). Observing time was also awarded with the ultra-wideband receiver, which spans 0.7 to 4 GHz; however, technical issues prevented data collection. Currently, observations are underway with the L-band receiver on the Green Bank Observatory's 20-meter Telescope, covering 1.3 to 1.8 GHz (5.4 to 7.4 $\mu$eV axion mass range), with data analysis in progress. In addition to M31, preliminary observations of the nearby young neutron star RBS1223 began in 2023 using the ARO 12-meter Telescope. Observations were carried out using 2 and 3 mm receivers, targeting the 84 to 95 GHz and 140 to 158 GHz ranges, corresponding to higher axion mass ranges (350 to 650 $\mu$eV). These observations are set to continue, with plans to apply the same analysis pipeline used for X-band data across these higher-frequency bands.

\section{Future goals}

Identifying the DM particle remains one of the central objectives of astrophysical research, and its discovery would have profound implications for both particle physics and astrophysics. Simulations suggest that, while individual AMCs are sparse, the cumulative effect of multiple axion structures in the Milky Way could produce a steady stream of faint signals, potentially detectable by future axion search experiments equipped with advanced radio technology. To advance this search, the establishment of a dedicated observing facility is proposed to continue the search for AMC-NS collisions and to determine the DM axion mass.
Given the rarity of these events, longer observation times are necessary to provide meaningful constraints on the axion parameter space, justifying the need for dedicated instrumentation. Should a candidate signal be detected, collaboration with laboratory experiments would be pursued to measure the axion-photon coupling constant $g_{a\gamma\gamma}$. Furthermore, both current and upcoming axion detection initiatives would benefit from a refined understanding of the distribution of overdensities, whose dynamics is critical to estimating the likelihood of observable events in laboratories.

\vspace{-0.2cm}
\begin{acknowledgments}
The authors thank their collaborators that helped with the development of the ideas presented in their papers: Prakamya Agrawal, Madeleine Edenton, Scott Ransom, Christoph Weniger, and Sam Witte. The authors thank the Green Bank Observatory for providing the observing time on the GBT (awards GBT22A-067 and GBT23A-245). LV acknowledges the support of the National Natural Science Foundation of China (NSFC) through Grant No.\ 12350610240 ``Astrophysical Axion Laboratories''. BJK acknowledges support from the DMpheno2lab Project (Project PID2022-139494NB-I00 financed by MCIN /AEI /10.13039/501100011033 / FEDER, UE). D.J.E.M.\ is supported by an Ernest Rutherford Fellowship from the STFC, Grant No.\ ST/T004037/1. This article is based on the work from COST Action COSMIC WISPers CA21106, supported by COST (European Cooperation in Science and Technology).
\end{acknowledgments}

{\footnotesize \bibliography{references.bib}}

\providecommand{\href}[2]{#2}\begingroup\raggedright\begin{thebibliography}{10}

\bibitem{Walters:2024vaw}
L.~Walters, J.~E. Shroyer, M.~Edenton, P.~Agrawal, B.~Johnson, B.~J. Kavanagh
  et~al., \emph{{Axions in Andromeda: Searching for minicluster-neutron star
  encounters with the Green Bank Telescope}},
  \href{https://doi.org/10.1103/PhysRevD.110.123002}{\emph{Phys. Rev. D}
  {\bfseries 110} (2024) 123002}
  [\href{https://arxiv.org/abs/2407.13060}{{\ttfamily 2407.13060}}].

\bibitem{Edwards:2020afl}
T.~D.~P. Edwards, B.~J. Kavanagh, L.~Visinelli and C.~Weniger, \emph{{Transient
  Radio Signatures from Neutron Star Encounters with QCD Axion Miniclusters}},
  \href{https://doi.org/10.1103/PhysRevLett.127.131103}{\emph{Phys. Rev. Lett.}
  {\bfseries 127} (2021) 131103}
  [\href{https://arxiv.org/abs/2011.05378}{{\ttfamily 2011.05378}}].

\bibitem{Kavanagh:2020gcy}
B.~J. Kavanagh, T.~D.~P. Edwards, L.~Visinelli and C.~Weniger, \emph{{Stellar
  disruption of axion miniclusters in the Milky~Way}},
  \href{https://doi.org/10.1103/PhysRevD.104.063038}{\emph{Phys. Rev. D}
  {\bfseries 104} (2021) 063038}
  [\href{https://arxiv.org/abs/2011.05377}{{\ttfamily 2011.05377}}].

\bibitem{Peccei:1977hh}
R.~D. Peccei and H.~R. Quinn, \emph{{CP Conservation in the Presence of
  Instantons}}, \href{https://doi.org/10.1103/PhysRevLett.38.1440}{\emph{Phys.\
  Rev.\ Lett.} {\bfseries 38} (1977) 1440}.

\bibitem{Peccei:1977ur}
R.~D. Peccei and H.~R. Quinn, \emph{{Constraints Imposed by CP Conservation in
  the Presence of Instantons}},
  \href{https://doi.org/10.1103/PhysRevD.16.1791}{\emph{Phys. Rev. D}
  {\bfseries 16} (1977) 1791}.

\bibitem{Weinberg:1977ma}
S.~Weinberg, \emph{{A New Light Boson?}},
  \href{https://doi.org/10.1103/PhysRevLett.40.223}{\emph{Phys. Rev. Lett.}
  {\bfseries 40} (1978) 223}.

\bibitem{Wilczek:1977pj}
F.~Wilczek, \emph{{Problem of Strong $P$ and $T$ Invariance in the Presence of
  Instantons}}, \href{https://doi.org/10.1103/PhysRevLett.40.279}{\emph{Phys.
  Rev. Lett.} {\bfseries 40} (1978) 279}.

\bibitem{Abbott:1982af}
L.~F. Abbott and P.~Sikivie, \emph{{A Cosmological Bound on the Invisible
  Axion}}, \href{https://doi.org/10.1016/0370-2693(83)90638-X}{\emph{Phys.
  Lett. B} {\bfseries 120} (1983) 133}.

\bibitem{Dine:1982ah}
M.~Dine and W.~Fischler, \emph{{The Not So Harmless Axion}},
  \href{https://doi.org/10.1016/0370-2693(83)90639-1}{\emph{Phys. Lett. B}
  {\bfseries 120} (1983) 137}.

\bibitem{Preskill:1982cy}
J.~Preskill, M.~B. Wise and F.~Wilczek, \emph{{Cosmology of the Invisible
  Axion}}, \href{https://doi.org/10.1016/0370-2693(83)90637-8}{\emph{Phys.
  Lett. B} {\bfseries 120} (1983) 127}.

\bibitem{Irastorza:2018dyq}
I.~G. Irastorza and J.~Redondo, \emph{{New experimental approaches in the
  search for axion-like particles}},
  \href{https://doi.org/10.1016/j.ppnp.2018.05.003}{\emph{Prog. Part. Nucl.
  Phys.} {\bfseries 102} (2018) 89}
  [\href{https://arxiv.org/abs/1801.08127}{{\ttfamily 1801.08127}}].

\bibitem{DiLuzio:2020wdo}
L.~Di~Luzio, M.~Giannotti, E.~Nardi and L.~Visinelli, \emph{{The landscape of
  QCD axion models}},
  \href{https://doi.org/10.1016/j.physrep.2020.06.002}{\emph{Phys. Rept.}
  {\bfseries 870} (2020) 1} [\href{https://arxiv.org/abs/2003.01100}{{\ttfamily
  2003.01100}}].

\bibitem{Chadha-Day:2021szb}
F.~Chadha-Day, J.~Ellis and D.~J.~E. Marsh, \emph{{Axion dark matter: What is
  it and why now?}}, \href{https://doi.org/10.1126/sciadv.abj3618}{\emph{Sci.
  Adv.} {\bfseries 8} (2022) abj3618}
  [\href{https://arxiv.org/abs/2105.01406}{{\ttfamily 2105.01406}}].

\bibitem{Giannotti:2024xhx}
M.~Giannotti, \emph{{Status and Perspectives on Axion Searches}},  12, 2024,
  \href{https://arxiv.org/abs/2412.08733}{{\ttfamily 2412.08733}}.

\bibitem{Sikivie:1983ip}
P.~Sikivie, \emph{{Experimental Tests of the Invisible Axion}},
  \href{https://doi.org/10.1103/PhysRevLett.51.1415}{\emph{Phys. Rev. Lett.}
  {\bfseries 51} (1983) 1415}.

\bibitem{Sikivie:1985yu}
P.~Sikivie, \emph{{Detection Rates for 'Invisible' Axion Searches}},
  \href{https://doi.org/10.1103/PhysRevD.36.974}{\emph{Phys. Rev. D} {\bfseries
  32} (1985) 2988}.

\bibitem{Wilczek:1987mv}
F.~Wilczek, \emph{{Two Applications of Axion Electrodynamics}},
  \href{https://doi.org/10.1103/PhysRevLett.58.1799}{\emph{Phys. Rev. Lett.}
  {\bfseries 58} (1987) 1799}.

\bibitem{Krasnikov:1996bm}
S.~V. Krasnikov, \emph{{New astrophysical constraints on the light pseudoscalar
  photon coupling}},
  \href{https://doi.org/10.1103/PhysRevLett.76.2633}{\emph{Phys. Rev. Lett.}
  {\bfseries 76} (1996) 2633}.

\bibitem{Visinelli:2013fia}
L.~Visinelli, \emph{{Axion-Electromagnetic Waves}},
  \href{https://doi.org/10.1142/S0217732313501629}{\emph{Mod. Phys. Lett. A}
  {\bfseries 28} (2013) 1350162}
  [\href{https://arxiv.org/abs/1401.0709}{{\ttfamily 1401.0709}}].

\bibitem{Tercas:2018gxv}
H.~Ter\c{c}as, J.~D. Rodrigues and J.~T. Mendon\c{c}a, \emph{{Axion-plasmon
  polaritons in strongly magnetized plasmas}},
  \href{https://doi.org/10.1103/PhysRevLett.120.181803}{\emph{Phys. Rev. Lett.}
  {\bfseries 120} (2018) 181803}
  [\href{https://arxiv.org/abs/1801.06254}{{\ttfamily 1801.06254}}].

\bibitem{Visinelli:2018zif}
L.~Visinelli and H.~Ter\c{c}as, \emph{{B-field induced mixing between Langmuir
  waves and axions}},
  \href{https://doi.org/10.1103/PhysRevD.105.096024}{\emph{Phys. Rev. D}
  {\bfseries 105} (2022) 096024}
  [\href{https://arxiv.org/abs/1807.06828}{{\ttfamily 1807.06828}}].

\bibitem{Hagmann:1998cb}
{\scshape ADMX} collaboration, \emph{{Results from a high sensitivity search
  for cosmic axions}},
  \href{https://doi.org/10.1103/PhysRevLett.80.2043}{\emph{Phys. Rev. Lett.}
  {\bfseries 80} (1998) 2043}
  [\href{https://arxiv.org/abs/astro-ph/9801286}{{\ttfamily
  astro-ph/9801286}}].

\bibitem{Asztalos:2001tf}
{\scshape ADMX} collaboration, \emph{{Large scale microwave cavity search for
  dark matter axions}},
  \href{https://doi.org/10.1103/PhysRevD.64.092003}{\emph{Phys. Rev. D}
  {\bfseries 64} (2001) 092003}.

\bibitem{Asztalos:2003px}
{\scshape ADMX} collaboration, \emph{{An Improved RF cavity search for halo
  axions}}, \href{https://doi.org/10.1103/PhysRevD.69.011101}{\emph{Phys. Rev.
  D} {\bfseries 69} (2004) 011101}
  [\href{https://arxiv.org/abs/astro-ph/0310042}{{\ttfamily
  astro-ph/0310042}}].

\bibitem{Graham:2015ouw}
P.~W. Graham, I.~G. Irastorza, S.~K. Lamoreaux, A.~Lindner and K.~A. van
  Bibber, \emph{{Experimental Searches for the Axion and Axion-Like
  Particles}},
  \href{https://doi.org/10.1146/annurev-nucl-102014-022120}{\emph{Ann. Rev.
  Nucl. Part. Sci.} {\bfseries 65} (2015) 485}
  [\href{https://arxiv.org/abs/1602.00039}{{\ttfamily 1602.00039}}].

\bibitem{Braine:2019fqb}
{\scshape ADMX} collaboration, \emph{{Extended Search for the Invisible Axion
  with the Axion Dark Matter Experiment}},
  \href{https://doi.org/10.1103/PhysRevLett.124.101303}{\emph{Phys. Rev. Lett.}
  {\bfseries 124} (2020) 101303}
  [\href{https://arxiv.org/abs/1910.08638}{{\ttfamily 1910.08638}}].

\bibitem{Kenany:2016tta}
S.~Al~Kenany et~al., \emph{{Design and operational experience of a microwave
  cavity axion detector for the 20\textendash{}100 $\mu$eV range}},
  \href{https://doi.org/10.1016/j.nima.2017.02.012}{\emph{Nucl. Instrum. Meth.
  A} {\bfseries 854} (2017) 11}
  [\href{https://arxiv.org/abs/1611.07123}{{\ttfamily 1611.07123}}].

\bibitem{Brubaker:2016ktl}
B.~M. Brubaker et~al., \emph{{First results from a microwave cavity axion
  search at 24 $\mu$eV}},
  \href{https://doi.org/10.1103/PhysRevLett.118.061302}{\emph{Phys. Rev. Lett.}
  {\bfseries 118} (2017) 061302}
  [\href{https://arxiv.org/abs/1610.02580}{{\ttfamily 1610.02580}}].

\bibitem{TheMADMAXWorkingGroup:2016hpc}
{\scshape MADMAX Working Group} collaboration, \emph{{Dielectric Haloscopes: A
  New Way to Detect Axion Dark Matter}},
  \href{https://doi.org/10.1103/PhysRevLett.118.091801}{\emph{Phys. Rev. Lett.}
  {\bfseries 118} (2017) 091801}
  [\href{https://arxiv.org/abs/1611.05865}{{\ttfamily 1611.05865}}].

\bibitem{Majorovits:2016yvk}
{\scshape MADMAX Working Group} collaboration, \emph{{MADMAX: A new Dark Matter
  Axion Search using a Dielectric Haloscope}},  in \emph{{12th Patras Workshop
  on Axions, WIMPs and WISPs}}, 11, 2016,
  \href{https://doi.org/10.3204/DESY-PROC-2009-03/Majorovits_Bela}{DOI}
  [\href{https://arxiv.org/abs/1611.04549}{{\ttfamily 1611.04549}}].

\bibitem{Alesini:2017ifp}
D.~Alesini, D.~Babusci, D.~Di~Gioacchino, C.~Gatti, G.~Lamanna and C.~Ligi,
  \emph{{The KLASH Proposal}},
  \href{https://arxiv.org/abs/1707.06010}{{\ttfamily 1707.06010}}.

\bibitem{Alesini:2019nzq}
D.~Alesini et~al., \emph{{KLASH Conceptual Design Report}},
  \href{https://arxiv.org/abs/1911.02427}{{\ttfamily 1911.02427}}.

\bibitem{Kahn:2016aff}
Y.~Kahn, B.~R. Safdi and J.~Thaler, \emph{{Broadband and Resonant Approaches to
  Axion Dark Matter Detection}},
  \href{https://doi.org/10.1103/PhysRevLett.117.141801}{\emph{Phys. Rev. Lett.}
  {\bfseries 117} (2016) 141801}
  [\href{https://arxiv.org/abs/1602.01086}{{\ttfamily 1602.01086}}].

\bibitem{Ouellet:2018beu}
J.~L. Ouellet et~al., \emph{{First Results from ABRACADABRA-10 cm: A Search for
  Sub-$\mu$eV Axion Dark Matter}},
  \href{https://doi.org/10.1103/PhysRevLett.122.121802}{\emph{Phys. Rev. Lett.}
  {\bfseries 122} (2019) 121802}
  [\href{https://arxiv.org/abs/1810.12257}{{\ttfamily 1810.12257}}].

\bibitem{Budker:2013hfa}
D.~Budker, P.~W. Graham, M.~Ledbetter, S.~Rajendran and A.~Sushkov,
  \emph{{Proposal for a Cosmic Axion Spin Precession Experiment (CASPEr)}},
  \href{https://doi.org/10.1103/PhysRevX.4.021030}{\emph{Phys. Rev. X}
  {\bfseries 4} (2014) 021030}
  [\href{https://arxiv.org/abs/1306.6089}{{\ttfamily 1306.6089}}].

\bibitem{Barbieri:2016vwg}
R.~Barbieri, C.~Braggio, G.~Carugno, C.~S. Gallo, A.~Lombardi, A.~Ortolan
  et~al., \emph{{Searching for galactic axions through magnetized media: the
  QUAX proposal}},
  \href{https://doi.org/10.1016/j.dark.2017.01.003}{\emph{Phys. Dark Univ.}
  {\bfseries 15} (2017) 135}
  [\href{https://arxiv.org/abs/1606.02201}{{\ttfamily 1606.02201}}].

\bibitem{Lee:2020cfj}
S.~Lee, S.~Ahn, J.~Choi, B.~R. Ko and Y.~K. Semertzidis, \emph{{Axion Dark
  Matter Search around 6.7 $\mu$eV}},
  \href{https://doi.org/10.1103/PhysRevLett.124.101802}{\emph{Phys. Rev. Lett.}
  {\bfseries 124} (2020) 101802}
  [\href{https://arxiv.org/abs/2001.05102}{{\ttfamily 2001.05102}}].

\bibitem{Alesini:2022lnp}
D.~Alesini et~al., \emph{{Search for Galactic axions with a high-Q dielectric
  cavity}}, \href{https://doi.org/10.1103/PhysRevD.106.052007}{\emph{Phys. Rev.
  D} {\bfseries 106} (2022) 052007}
  [\href{https://arxiv.org/abs/2208.12670}{{\ttfamily 2208.12670}}].

\bibitem{Alesini:2023qed}
D.~Alesini et~al., \emph{{The future search for low-frequency axions and new
  physics with the FLASH resonant cavity experiment at Frascati National
  Laboratories}}, \href{https://doi.org/10.1016/j.dark.2023.101370}{\emph{Phys.
  Dark Univ.} {\bfseries 42} (2023) 101370}
  [\href{https://arxiv.org/abs/2309.00351}{{\ttfamily 2309.00351}}].

\bibitem{Gatti:2024mde}
C.~Gatti, L.~Visinelli and M.~Zantedeschi, \emph{{Cavity detection of
  gravitational waves: Where do we stand?}},
  \href{https://doi.org/10.1103/PhysRevD.110.023018}{\emph{Phys. Rev. D}
  {\bfseries 110} (2024) 023018}
  [\href{https://arxiv.org/abs/2403.18610}{{\ttfamily 2403.18610}}].

\bibitem{ADMX:2024pxg}
{\scshape ADMX} collaboration, \emph{{Search for non-virialized axions with
  3.3-4.2 $\mu$eV mass at selected resolving powers}},
  \href{https://arxiv.org/abs/2410.09203}{{\ttfamily 2410.09203}}.

\bibitem{OShea:2023gqn}
T.~O'Shea, M.~Giannotti, I.~G. Irastorza, L.~M. Plasencia, J.~Redondo, J.~Ruz
  et~al., \emph{{Prospects on the detection of solar dark photons by the
  International Axion Observatory}},
  \href{https://doi.org/10.1088/1475-7516/2024/06/070}{\emph{JCAP} {\bfseries
  06} (2024) 070} [\href{https://arxiv.org/abs/2312.10150}{{\ttfamily
  2312.10150}}].

\bibitem{OShea:2024jjw}
T.~O'Shea, A.-C. Davis, M.~Giannotti, S.~Vagnozzi, L.~Visinelli and J.~K.
  Vogel, \emph{{Solar chameleons: Novel channels}},
  \href{https://doi.org/10.1103/PhysRevD.110.063027}{\emph{Phys. Rev. D}
  {\bfseries 110} (2024) 063027}
  [\href{https://arxiv.org/abs/2406.01691}{{\ttfamily 2406.01691}}].

\bibitem{Ruz:2024gkl}
J.~Ruz et~al., \emph{{NuSTAR as an Axion Helioscope}},
  \href{https://arxiv.org/abs/2407.03828}{{\ttfamily 2407.03828}}.

\bibitem{Pshirkov:2007st}
M.~S. Pshirkov and S.~B. Popov, \emph{{Conversion of Dark matter axions to
  photons in magnetospheres of neutron stars}},
  \href{https://doi.org/10.1134/S1063776109030030}{\emph{J. Exp. Theor. Phys.}
  {\bfseries 108} (2009) 384}
  [\href{https://arxiv.org/abs/0711.1264}{{\ttfamily 0711.1264}}].

\bibitem{Huang:2018lxq}
F.~P. Huang, K.~Kadota, T.~Sekiguchi and H.~Tashiro, \emph{{Radio telescope
  search for the resonant conversion of cold dark matter axions from the
  magnetized astrophysical sources}},
  \href{https://doi.org/10.1103/PhysRevD.97.123001}{\emph{Phys. Rev. D}
  {\bfseries 97} (2018) 123001}
  [\href{https://arxiv.org/abs/1803.08230}{{\ttfamily 1803.08230}}].

\bibitem{Hook:2018iia}
A.~Hook, Y.~Kahn, B.~R. Safdi and Z.~Sun, \emph{{Radio Signals from Axion Dark
  Matter Conversion in Neutron Star Magnetospheres}},
  \href{https://doi.org/10.1103/PhysRevLett.121.241102}{\emph{Phys. Rev. Lett.}
  {\bfseries 121} (2018) 241102}
  [\href{https://arxiv.org/abs/1804.03145}{{\ttfamily 1804.03145}}].

\bibitem{Safdi:2018oeu}
B.~R. Safdi, Z.~Sun and A.~Y. Chen, \emph{{Detecting Axion Dark Matter with
  Radio Lines from Neutron Star Populations}},
  \href{https://doi.org/10.1103/PhysRevD.99.123021}{\emph{Phys. Rev. D}
  {\bfseries 99} (2019) 123021}
  [\href{https://arxiv.org/abs/1811.01020}{{\ttfamily 1811.01020}}].

\bibitem{Leroy:2019ghm}
M.~Leroy, M.~Chianese, T.~D.~P. Edwards and C.~Weniger, \emph{{Radio Signal of
  Axion-Photon Conversion in Neutron Stars: A Ray Tracing Analysis}},
  \href{https://doi.org/10.1103/PhysRevD.101.123003}{\emph{Phys. Rev. D}
  {\bfseries 101} (2020) 123003}
  [\href{https://arxiv.org/abs/1912.08815}{{\ttfamily 1912.08815}}].

\bibitem{Hogan:1988mp}
C.~J. Hogan and M.~J. Rees, \emph{{AXION MINICLUSTERS}},
  \href{https://doi.org/10.1016/0370-2693(88)91655-3}{\emph{Phys. Lett. B}
  {\bfseries 205} (1988) 228}.

\bibitem{Kolb:1993hw}
E.~W. Kolb and I.~I. Tkachev, \emph{{Nonlinear axion dynamics and formation of
  cosmological pseudosolitons}},
  \href{https://doi.org/10.1103/PhysRevD.49.5040}{\emph{Phys. Rev. D}
  {\bfseries 49} (1994) 5040}
  [\href{https://arxiv.org/abs/astro-ph/9311037}{{\ttfamily
  astro-ph/9311037}}].

\bibitem{Kolb:1993zz}
E.~W. Kolb and I.~I. Tkachev, \emph{{Axion miniclusters and Bose stars}},
  \href{https://doi.org/10.1103/PhysRevLett.71.3051}{\emph{Phys. Rev. Lett.}
  {\bfseries 71} (1993) 3051}
  [\href{https://arxiv.org/abs/hep-ph/9303313}{{\ttfamily hep-ph/9303313}}].

\bibitem{Zurek:2006sy}
K.~M. Zurek, C.~J. Hogan and T.~R. Quinn, \emph{{Astrophysical Effects of
  Scalar Dark Matter Miniclusters}},
  \href{https://doi.org/10.1103/PhysRevD.75.043511}{\emph{Phys. Rev. D}
  {\bfseries 75} (2007) 043511}
  [\href{https://arxiv.org/abs/astro-ph/0607341}{{\ttfamily
  astro-ph/0607341}}].

\bibitem{Vaquero:2018tib}
A.~Vaquero, J.~Redondo and J.~Stadler, \emph{{Early seeds of axion
  miniclusters}},
  \href{https://doi.org/10.1088/1475-7516/2019/04/012}{\emph{JCAP} {\bfseries
  04} (2019) 012} [\href{https://arxiv.org/abs/1809.09241}{{\ttfamily
  1809.09241}}].

\bibitem{Buschmann:2019icd}
M.~Buschmann, J.~W. Foster and B.~R. Safdi, \emph{{Early-Universe Simulations
  of the Cosmological Axion}},
  \href{https://doi.org/10.1103/PhysRevLett.124.161103}{\emph{Phys. Rev. Lett.}
  {\bfseries 124} (2020) 161103}
  [\href{https://arxiv.org/abs/1906.00967}{{\ttfamily 1906.00967}}].

\bibitem{Sikivie:2006ni}
P.~Sikivie, \emph{{Axion Cosmology}},
  \href{https://doi.org/10.1007/978-3-540-73518-2_2}{\emph{Lect. Notes Phys.}
  {\bfseries 741} (2008) 19}
  [\href{https://arxiv.org/abs/astro-ph/0610440}{{\ttfamily
  astro-ph/0610440}}].

\bibitem{Hertzberg:2020hsz}
M.~P. Hertzberg, E.~D. Schiappacasse and T.~T. Yanagida, \emph{{Axion Star
  Nucleation in Dark Minihalos around Primordial Black Holes}},
  \href{https://doi.org/10.1103/PhysRevD.102.023013}{\emph{Phys. Rev. D}
  {\bfseries 102} (2020) 023013}
  [\href{https://arxiv.org/abs/2001.07476}{{\ttfamily 2001.07476}}].

\bibitem{Yin:2024xov}
Z.~Yin and L.~Visinelli, \emph{{Axion star condensation around primordial black
  holes and microlensing limits}},
  \href{https://doi.org/10.1088/1475-7516/2024/10/013}{\emph{JCAP} {\bfseries
  10} (2024) 013} [\href{https://arxiv.org/abs/2404.10340}{{\ttfamily
  2404.10340}}].

\bibitem{Berezinsky:2013fxa}
V.~S. Berezinsky, V.~I. Dokuchaev and Y.~N. Eroshenko, \emph{{Formation and
  internal structure of superdense dark matter clumps and ultracompact
  minihaloes}},
  \href{https://doi.org/10.1088/1475-7516/2013/11/059}{\emph{JCAP} {\bfseries
  11} (2013) 059} [\href{https://arxiv.org/abs/1308.6742}{{\ttfamily
  1308.6742}}].

\bibitem{Tinyakov:2015cgg}
P.~Tinyakov, I.~Tkachev and K.~Zioutas, \emph{{Tidal streams from axion
  miniclusters and direct axion searches}},
  \href{https://doi.org/10.1088/1475-7516/2016/01/035}{\emph{JCAP} {\bfseries
  01} (2016) 035} [\href{https://arxiv.org/abs/1512.02884}{{\ttfamily
  1512.02884}}].

\bibitem{Dandoy:2022prp}
V.~Dandoy, T.~Schwetz and E.~Todarello, \emph{{A self-consistent wave
  description of axion miniclusters and their survival in the galaxy}},
  \href{https://doi.org/10.1088/1475-7516/2022/09/081}{\emph{JCAP} {\bfseries
  09} (2022) 081} [\href{https://arxiv.org/abs/2206.04619}{{\ttfamily
  2206.04619}}].

\bibitem{Shen:2022ltx}
X.~Shen, H.~Xiao, P.~F. Hopkins and K.~M. Zurek, \emph{{Disruption of Dark
  Matter Minihalos in the Milky Way Environment: Implications for Axion
  Miniclusters and Early Matter Domination}},
  \href{https://doi.org/10.3847/1538-4357/ad12c6}{\emph{Astrophys. J.}
  {\bfseries 962} (2024) 9} [\href{https://arxiv.org/abs/2207.11276}{{\ttfamily
  2207.11276}}].

\bibitem{OHare:2023rtm}
C.~A.~J. O'Hare, G.~Pierobon and J.~Redondo, \emph{{Axion Minicluster Streams
  in the Solar Neighborhood}},
  \href{https://doi.org/10.1103/PhysRevLett.133.081001}{\emph{Phys. Rev. Lett.}
  {\bfseries 133} (2024) 081001}
  [\href{https://arxiv.org/abs/2311.17367}{{\ttfamily 2311.17367}}].

\bibitem{DSouza:2024flu}
I.~DSouza and C.~Gordon, \emph{{Disruption of dark matter minihalos by
  successive stellar encounters}},
  \href{https://doi.org/10.1103/PhysRevD.109.123035}{\emph{Phys. Rev. D}
  {\bfseries 109} (2024) 123035}
  [\href{https://arxiv.org/abs/2402.03236}{{\ttfamily 2402.03236}}].

\bibitem{Levkov:2018kau}
D.~G. Levkov, A.~G. Panin and I.~I. Tkachev, \emph{{Gravitational Bose-Einstein
  condensation in the kinetic regime}},
  \href{https://doi.org/10.1103/PhysRevLett.121.151301}{\emph{Phys. Rev. Lett.}
  {\bfseries 121} (2018) 151301}
  [\href{https://arxiv.org/abs/1804.05857}{{\ttfamily 1804.05857}}].

\bibitem{Eggemeier:2019jsu}
B.~Eggemeier and J.~C. Niemeyer, \emph{{Formation and mass growth of axion
  stars in axion miniclusters}},
  \href{https://doi.org/10.1103/PhysRevD.100.063528}{\emph{Phys. Rev. D}
  {\bfseries 100} (2019) 063528}
  [\href{https://arxiv.org/abs/1906.01348}{{\ttfamily 1906.01348}}].

\bibitem{Chen:2020cef}
J.~Chen, X.~Du, E.~W. Lentz, D.~J.~E. Marsh and J.~C. Niemeyer, \emph{{New
  insights into the formation and growth of boson stars in dark matter halos}},
  \href{https://doi.org/10.1103/PhysRevD.104.083022}{\emph{Phys. Rev. D}
  {\bfseries 104} (2021) 083022}
  [\href{https://arxiv.org/abs/2011.01333}{{\ttfamily 2011.01333}}].

\bibitem{Dmitriev:2023ipv}
A.~S. Dmitriev, D.~G. Levkov, A.~G. Panin and I.~I. Tkachev,
  \emph{{Self-Similar Growth of Bose Stars}},
  \href{https://doi.org/10.1103/PhysRevLett.132.091001}{\emph{Phys. Rev. Lett.}
  {\bfseries 132} (2024) 091001}
  [\href{https://arxiv.org/abs/2305.01005}{{\ttfamily 2305.01005}}].

\bibitem{Visinelli:2017ooc}
L.~Visinelli, S.~Baum, J.~Redondo, K.~Freese and F.~Wilczek, \emph{{Dilute and
  dense axion stars}},
  \href{https://doi.org/10.1016/j.physletb.2017.12.010}{\emph{Phys. Lett. B}
  {\bfseries 777} (2018) 64}
  [\href{https://arxiv.org/abs/1710.08910}{{\ttfamily 1710.08910}}].

\bibitem{Schiappacasse:2017ham}
E.~D. Schiappacasse and M.~P. Hertzberg, \emph{{Analysis of Dark Matter Axion
  Clumps with Spherical Symmetry}},
  \href{https://doi.org/10.1088/1475-7516/2018/01/037}{\emph{JCAP} {\bfseries
  01} (2018) 037} [\href{https://arxiv.org/abs/1710.04729}{{\ttfamily
  1710.04729}}].

\bibitem{Visinelli:2009kt}
L.~Visinelli and P.~Gondolo, \emph{{Axion cold dark matter in non-standard
  cosmologies}}, \href{https://doi.org/10.1103/PhysRevD.81.063508}{\emph{Phys.
  Rev. D} {\bfseries 81} (2010) 063508}
  [\href{https://arxiv.org/abs/0912.0015}{{\ttfamily 0912.0015}}].

\bibitem{Eggemeier:2019khm}
B.~Eggemeier, J.~Redondo, K.~Dolag, J.~C. Niemeyer and A.~Vaquero, \emph{{First
  Simulations of Axion Minicluster Halos}},
  \href{https://doi.org/10.1103/PhysRevLett.125.041301}{\emph{Phys. Rev. Lett.}
  {\bfseries 125} (2020) 041301}
  [\href{https://arxiv.org/abs/1911.09417}{{\ttfamily 1911.09417}}].

\bibitem{Eggemeier:2022hqa}
B.~Eggemeier, C.~A.~J. O'Hare, G.~Pierobon, J.~Redondo and Y.~Y.~Y. Wong,
  \emph{{Axion minivoids and implications for direct detection}},
  \href{https://doi.org/10.1103/PhysRevD.107.083510}{\emph{Phys. Rev. D}
  {\bfseries 107} (2023) 083510}
  [\href{https://arxiv.org/abs/2212.00560}{{\ttfamily 2212.00560}}].

\bibitem{Buschmann:2021sdq}
M.~Buschmann, J.~W. Foster, A.~Hook, A.~Peterson, D.~E. Willcox, W.~Zhang
  et~al., \emph{{Dark matter from axion strings with adaptive mesh
  refinement}}, \href{https://doi.org/10.1038/s41467-022-28669-y}{\emph{Nature
  Commun.} {\bfseries 13} (2022) 1049}
  [\href{https://arxiv.org/abs/2108.05368}{{\ttfamily 2108.05368}}].

\bibitem{Navarro:1995iw}
J.~F. Navarro, C.~S. Frenk and S.~D.~M. White, \emph{{The Structure of cold
  dark matter halos}}, \href{https://doi.org/10.1086/177173}{\emph{Astrophys.
  J.} {\bfseries 462} (1996) 563}
  [\href{https://arxiv.org/abs/astro-ph/9508025}{{\ttfamily
  astro-ph/9508025}}].

\bibitem{Dokuchaev:2017psd}
V.~I. Dokuchaev, Y.~N. Eroshenko and I.~I. Tkachev, \emph{{Destruction of axion
  miniclusters in the Galaxy}},
  \href{https://doi.org/10.1134/S1063776117080039}{\emph{J. Exp. Theor. Phys.}
  {\bfseries 125} (2017) 434}
  [\href{https://arxiv.org/abs/1710.09586}{{\ttfamily 1710.09586}}].

\bibitem{Witte:2021arp}
S.~J. Witte, D.~Noordhuis, T.~D.~P. Edwards and C.~Weniger, \emph{{Axion-photon
  conversion in neutron star magnetospheres: The role of the plasma in the
  Goldreich-Julian model}},
  \href{https://doi.org/10.1103/PhysRevD.104.103030}{\emph{Phys. Rev. D}
  {\bfseries 104} (2021) 103030}
  [\href{https://arxiv.org/abs/2104.07670}{{\ttfamily 2104.07670}}].

\bibitem{OHare:2021zrq}
C.~A.~J. O'Hare, G.~Pierobon, J.~Redondo and Y.~Y.~Y. Wong, \emph{{Simulations
  of axionlike particles in the postinflationary scenario}},
  \href{https://doi.org/10.1103/PhysRevD.105.055025}{\emph{Phys. Rev. D}
  {\bfseries 105} (2022) 055025}
  [\href{https://arxiv.org/abs/2112.05117}{{\ttfamily 2112.05117}}].

\bibitem{Gorghetto:2022ikz}
M.~Gorghetto and E.~Hardy, \emph{{Post-inflationary axions: a minimal target
  for axion haloscopes}},
  \href{https://doi.org/10.1007/JHEP05(2023)030}{\emph{JHEP} {\bfseries 05}
  (2023) 030} [\href{https://arxiv.org/abs/2212.13263}{{\ttfamily
  2212.13263}}].

\end{thebibliography}\endgroup

\end{document}